\documentclass{ieeeaccess}
\makeatletter
\let\MYcaption\@makecaption
\makeatother
\usepackage{subcaption}
\makeatletter
\let\@makecaption\MYcaption
\makeatother
\usepackage{cite}
\usepackage{dcolumn}
\usepackage{bm}
\usepackage{amsmath,amssymb,amsfonts}
\usepackage{enumitem}
\usepackage{physics}
\usepackage{multirow}
\usepackage{graphicx}
\usepackage{blkarray}
\usepackage{xspace}
\usepackage{tabularx}
\usepackage[T1,T2A]{fontenc}
\usepackage{siunitx}
\usepackage{balance}
\usepackage[normalem]{ulem}
\usepackage{xspace}

\newcommand{\Conf}{{\bf Confidentiality}\xspace}
\newcommand{\Avail}{{\bf Availability}\xspace}
\newcommand{\Integ}{{\bf Integrity}\xspace}

\def\BibTeX{{\rm B\kern-.05em{\sc i\kern-.025em b}\kern-.08em
    T\kern-.1667em\lower.7ex\hbox{E}\kern-.125emX}}
\begin{document}
\history{Date of publication xxxx 00, 0000, date of current version xxxx 00, 0000.}
\doi{10.1109/TQE.2020.DOI}

\title{Attacking the Quantum Internet}
\author{\uppercase{Takahiko Satoh}\authorrefmark{1,2},
\uppercase{Shota Nagayama}\authorrefmark{3}, \IEEEmembership{Member, IEEE}, 
\uppercase{Shigeya Suzuki}\authorrefmark{4}, \IEEEmembership{Member, IEEE},
\uppercase{Takaaki Matsuo}\authorrefmark{4},
\uppercase{Michal Hajdu\v{s}ek\authorrefmark{4}, and Rodney Van Meter}.\authorrefmark{1,5}, \IEEEmembership{Senior Member, IEEE}}
\address[1]{Keio University Quantum Computing Center, Yokohama, Kanagawa 223-8522 Japan}
\address[2]{Graduate School of Science and Technology, Keio University, Yokohama, Kanagawa 223-8522 Japan}
\address[3]{Mercari, Inc.}
\address[4]{Graduate  School  of  Media  and  Governance, Keio University SFC, Fujisawa, 
Kanagawa 252-0882 Japan}
\address[5]{Faculty of Environment and Information Studies, Keio University SFC, Fujisawa, 
Kanagawa 252-0882 Japan}
\tfootnote{This material is based upon work supported by the Air Force Office of Scientific Research under award number FA2386-19-1-4038. Portions of this paper appeared in an NDSS workshop paper by S. Suzuki and R. Van Meter~\cite{suzuki2015classification}. This paper also extends work from Satoh {\it et al.}~\cite{satoh2018network}.}

\markboth
{Satoh \headeretal: Attacking the Quantum Internet}
{Satoh \headeretal: Attacking the Quantum Internet}

\corresp{Corresponding author: Takahiko Satoh (email: satoh@sfc.wide.ad.jp).}

\begin{abstract}
The main service provided by the coming Quantum Internet will be
creating entanglement between any two quantum nodes.  We discuss and
classify attacks on quantum repeaters, which will serve 
roles similar to those of classical Internet routers.
We have modeled the components for and structure of quantum repeater
network nodes.  With this model, we point out attack vectors, then
analyze attacks in terms of confidentiality, integrity and
availability.  While we are reassured about the promises of quantum
networks from the confidentiality point of view, integrity and availability present new vulnerabilities not present in classical networks and require care to handle properly.  We observe that the
requirements on the classical computing/networking elements affect the
systems' overall security risks.  This component-based analysis
establishes a framework for further investigation of network-wide
vulnerabilities.
\end{abstract}


\titlepgskip=-15pt
\maketitle
\section{Introduction}\label{sec:intro}
\IEEEPARstart{T}{he}  computers and networks in common use today are built on classical notions of information, generally using small amounts of electrical charge, the orientation of tiny magnets, and optical signals as data. 
We typically treat the data states as binary numbers or symbols and manipulate them using familiar, comfortable Boolean logic. 
But over the last three decades, a new theory of information based on quantum mechanics has been discovered, quantum algorithms have been developed, experimental demonstrations of quantum computing have proliferated, and large-scale machines are on the drawing boards~\cite{bacon2010recent,montanaro2015:qualgo-qi,NC,van-meter2013blueprint,wilde2013quantum}.
One of the oldest and most successful areas in quantum information has been quantum networks~\cite{dahlberg2019link,kimble08:_quant_internet,van-meter14,Wehner18:eaam9288,PhysRevLett.120.030503}.

Work on quantum networks began with the recognition that transmitted qubits act as exquisite sensors of state modification, and can be used to detect the presence of eavesdroppers on a quantum communication channel while creating shared, secret
random numbers useful as keys for encrypting classical data, known as \emph{quantum key
  distribution}~(QKD~\cite{bb84,ekert1991qcb}).  The array of proposed applications for distributed quantum information has grown
to include other cybernetic uses such as clock synchronization, reference frame alignment, and interferometry for
astronomy~\cite{RevModPhys.79.555,PhysRevLett.109.070503,jozsa2000qcs}.  
Distributed quantum computation will help to build large scale quantum computers, especially by combining heterogeneous quantum modules~\cite{PhysRevA.59.4249,grover1997quantum,nagayama17:thesis,doi:10.1002/1521-3978(200009)48:9/11<839::AID-PROP839>3.0.CO;2-V, 10.1145/1150019.1136517}.
The development of large-scale quantum
computers would affect classical security systems that depend on the difficulty of certain computational problems, but
conversely distributed security-related functions such as Byzantine agreement and secret sharing recoup some of those
losses~\cite{byzantine,crepeau:_secur_multi_party_qc}.  Broadbent {\it et al}. developed a fully blind
method of conducting any arbitrary quantum calculation~(BQC~\cite{broadbent2010measurement,chien15:_ft-blind}).
Unlike
Gentry's classical homomorphic encryption~\cite{Gentry:2010:CAF:1666420.1666444}, this technique hides the algorithm
itself as well as the input and output data.  Thus, if we can find ways of distributing quantum information over long
distances, we will enable valuable new functionality.

\emph{Quantum entanglement} is a correlation between the states of two or more qubits, stronger than any possible classical correlation~\cite{horodecki2009quantum,werner2001bell}.
Although entanglement cannot be used to transmit information faster than the speed of light, two qubits may be in an entangled state where their values are decided randomly but seemingly in an instantaneously coordinated fashion \emph{without} any apparent communication.
Many of the applications just discussed require us to create this entanglement over a distance. 
\emph{Quantum repeaters} (Sec.~\ref{sec:qrintro}) are an important path toward building a Quantum Internet that will achieve this goal by using quantum entanglement for quantum teleportation.

The early use of the Quantum Internet with high noise levels would be to enhance Internet security with QKD~\cite{PhysRevLett.77.2818,elliott02:_building_quant_network,elliott:qkd-net}.
Various attacks for preventing such security improvement have already been proposed~\cite{gerhardt2011full,jogenfors2015hacking,lydersen2010hacking,makarov2005faked,pirker17:security}.
Defense methods, operational methods with optimal efficiency, and specific methods for combining with classical protocols have also been proposed~\cite{PhysRevLett.98.230501,bratzik13:_quant_repeaters_and_qkd,mink09:_qkd_and_ipsec}.
Urban-scale networks have already been built by trusting intermediate nodes to avoid the requirement of quantum repeaters, and their performance has been demonstrated~\cite{peev:secoqc, sasaki2011field, 1367-2630-13-12-123001, wang2014field}.

The classical Internet, the global-scale network of networks, has emerged over some five decades, and security is a major area in research, engineering and operations~\cite{bellovin1989security,bishop2002art}.
Both hardware and software evolve quickly, and both attacks and defense applied to network infrastructure and end nodes emerge at an astounding rate. 
Some attacks compromise individual computers or data, either during the initiation or data transfer phases of a communication session, by spoofing data packets, hijacking connections, or cracking encryption. 
Attacks on sessions can also be attempted more speculatively by compromising systems, then laying in wait for opportunities to present themselves.
Other vulnerabilities affect the stability of the network itself by disrupting routing or naming systems, or by flooding portions of the network with excess traffic.
Such vulnerabilities and attacks have to be discussed to design secure Quantum Internet architectures.

In this paper, we summarize and develop primitive models of attacks on individual components of quantum networks.
We do this within the context of \Conf, \Integ and \Avail, often abbreviated as the CIA triad.
\Conf of a quantum network ensures that no information is leaked to an unauthorized party.
\Integ ensures that the quantum information is accurate and trustworthy.
\Avail guarantees access to the quantum information by authorized parties.
These definitions are identical to their usual meaning in classical cybersecurity, however their interaction may be different in quantum networks due to the presence of entanglement.

As an example, consider the case of quantum teleportation~\cite{teleportation} where a sender wishes to communicate a single quantum bit to a receiver with the use of a shared entangled pair and two classical bits.
Let's assume an unauthorized third party manages to both steal one half of the entangled pair intended for the receiver without being detected by quantum state certification and gain access to the classical channel.
In this case, \Conf of the quantum state is compromised because the unauthorized party can use the two classical bits to successfully complete the teleportation protocol.
On the other hand, if the classical channel remains uncompromised, the unauthorized party cannot complete the teleportation protocol and therefore does not affect \Conf.
In fact, without the two classical bits the unauthorized party is in possession of a maximally mixed state while the original quantum state is destroyed on the sender's end.
In this case, it is the \Integ of the quantum information that is affected.
Depending on the type of attack on the quantum network, one or multiple aspects of the CIA triad may be compromised.
We will see that particularly \Conf and \Integ of quantum information are very closely related due to how quantum information is communicated in quantum networks.

The attackers' purpose may be parallel to those in classical networks:
\begin{itemize}
\item {\it to steal} quantum information; or
\item {\it to disrupt} either the \Integ or \Avail of quantum nodes or quantum networks; or
\item {\it to hijack} a quantum connection or computing resources such as control of quantum repeaters or external components.
\end{itemize}

The biggest difference between classical and quantum networks is the presence of entanglement.
This difference raises questions:
\begin{itemize}
\item Can an attacker copy or disclose the transferred quantum data via an illicitly created entanglement?
\item Can an attacker compromise later sessions by hijacking qubits or by undetected malicious acts?
\end{itemize}
Even without entanglement, new questions are raised, such as
\begin{itemize}
\item Can control of the quantum hardware elements allow hijacking of the repeater itself or disclosure of the stored information? 
\item Classical hardware is vulnerable to damage from strong electrical or optical pulses. Are quantum nodes more vulnerable than classical systems? (This question is dependent on implementation, and is a moving target we will not address here.)
\end{itemize}
More generally, to assess the scalability and stability of the  Quantum Internet, 
\begin{itemize}
\item can the function of creating end-to-end entanglement be disrupted on a scale disproportionate to the fraction of the network compromise?
\end{itemize}
While attacks attempting theft target operations of a communication session, this question conjures effects on fundamental network functions such as routing~\cite{van-meter:qDijkstra}.
The process of answering these questions will certainly last far into the future.
This paper gives the first framework to categorize attacks in pursuit of these goals.

RFID systems exchange information by letting short-range wireless communication act on an RF tag with ID information.
We noted that RFID systems and quantum repeater systems are similar in that they are hybrid systems with tightly coupled sensing and software elements.
Since RFID systems are sensitive to noise and intentional, malicious input, a classification of attack methods, such as information theft and spoofing, has been developed by Weingart~\cite{weingart2000psd}, Mitrokotsa~\cite{mitrokotsa2010ttn}, and Mirowski~\cite{mirowski2009ara}.
These methods inspired us to classify attacks on quantum networks.

While we can model the basic hardware architecture of a quantum network nodes and have some idea of required elements, a concrete design for a specific implementation of such a system has not been achieved yet.
We begin with an overview of the Quantum Internet (Sec.~\ref{sec:qrintro})
and a hardware model that will allow us to identify points of attack, then classify the primitive attacks (Sec.~\ref{sec:model}).
We then investigate the means of attack on the Quantum Internet through the elements of the Quantum Nodes (Sec.~\ref{sec:primitive}), and also discuss what an attacker who has hijacked control of one or more Quantum Nodes can do (Sec.~\ref{sec:pl}).
We believe that this paper will contribute toward designing secure Quantum Internet architectures.
In such work, knowledge gained during the engineering of classical networks will be beneficial to minimize security issues of developing quantum networks.

\section{Quantum Internet}\label{sec:qrintro}
We have already introduced the concept of quantum entanglement and what it is good for, but not how widely distributed entanglement can be created. 
A network of optical links connected by \emph{quantum repeaters} will fill the role of classical network links and switches or routers.
End nodes that can connect to the quantum network will provide various quantum services that enable the uses discussed above.
As in the classical Internet, individual quantum networks of potentially heterogeneous technology and independent management will ultimately come together to form a Quantum Internet~\cite{van-meter14}.

\subsection{The role of a Quantum Repeater}
To perform long distance communication, a quantum repeater must supply the following four functions.
\subsubsection{Node-to-Node Entanglement generation}
Experimental physicists have demonstrated the creation of entanglement over short distances using single photons (e.g.~\cite{moehring2007eos}).
Numerous approaches have been proposed and some of them demonstrated, but for our purposes here a single example will suffice.
Individual quantum bits, or \emph{qubits}, at each node may be single atoms suspended in a vacuum or another of the dozens of technologies under experimental development~\cite{ladd2010quantum,van-meter2013blueprint}.
A qubit at each end of a link is coaxed to emit a photon that is entangled with the qubit. 
The two photons are routed toward each other and ultimately interfere in a fashion that erases knowledge of where each photon came from, leaving the two stationary qubits entangled in what is called a \emph{Bell pair}, named for a proposal made by John Bell over fifty years ago~\cite{bell1964ote}.

\subsubsection{Stretching of Entanglement}
Naturally, we can't transmit those photons over arbitrary distances.  
In optical fibers, the probability of success falls exponentially with distance as photons are lost.
Unfortunately, in quantum networks, classical amplifiers cannot be used because independent copies of quantum data cannot be made due to the no-cloning theorem~\cite{no_cloning}.

Moreover, in any interesting network, we want to support multi-hop paths between pairs of nodes, rather than requiring a direct link between each pair.
Both problems can be solved by using \emph{entanglement swapping}, which takes two Bell pairs, one between nodes $A$ and $B$ and one between nodes $B$ and $C$, and splices them together to form a single Bell pair that spans from $A$ to $C$~\cite{PhysRevLett.71.4287}.  Entanglement swapping can be viewed as an extension of teleportation to entangled states.

\subsubsection{Management of errors}
The quality, or fidelity (a measure between two quantum states, generally, ideal state and actual state, hence it describes the quality of qubits), of these Bell pairs declines as we perform more of these swapping operations, eventually destroying the quantumness of the data and leaving only random classical noise.
This problem can be solved by using a form of error detection known as \emph{purification}~\cite{bennett1996purification} or using quantum error correction~\cite{devitt2013quantum,Briegel_2007}.
Purification plus entanglement swapping is the canonical setup of a chain of quantum repeaters~\cite{briegel98:_quant_repeater}.

\subsubsection{Management of the network}
Nodes must also participate actively in the management of the network itself, including routing and multiplexing or resource management.  This work is the focus of a number of attacks described in this paper.

\subsection{Types of nodes}
\label{sec:nodetype}
We can classify quantum network nodes (QNodes) under five types by the number of
connected (quantum) links and their roles:

\subsubsection*{End node (ENode)}
An ENode works as a terminal node for running quantum applications. 
In order to support various applications, qubit operations and memory functions are required.
An ENode has exactly one external link and corresponds to clients and servers.

\subsubsection*{Measurement node (MNode)}
An MNode is a terminal node just for measurement, used by end users for QKD or blind quantum computation~\cite{PhysRevA.89.060302,PhysRevA.87.050301}.
The only necessary function is to measure qubits,
therefore an MNode has no static memory.
An MNode has exactly one external link.
We can regard an MNode as a simpler ENode.

\subsubsection*{Repeater node (RNode)} 
RNodes are installed at fixed distances according to the optical fiber loss level to improve network performance in long-distance quantum communications~\cite{briegel98:_quant_repeater}.
An RNode has exactly two external links, so that it is useful in a line only.
RNodes correspond to repeaters in classical networking.

\subsubsection*{Router (XNode)} 
An XNode has two or more external links, e.g., connected internally via an optical backplane~\cite{nagayama17:thesis}.
XNodes have more substantial processing capacities than other node types and are responsible for branching the route of the network in complex network topologies.  They also serve as network boundaries in an internetwork.

\subsubsection*{Intermediate node (INode)}
INodes may be placed between the above QNodes and are responsible for generating and connecting Bell pairs.
An INode has exactly two external links.
Whether a link between QNodes contains an INode or not depends on the Bell pair sharing scheme.
There are several configurations of INodes and link, and we describe the elements and functions of INodes in Sec.~\ref{subsec:typeqn}. 
Unless otherwise noted, we do {\bf not} include INode in our discussion of QNode.

\vskip 1.0em

Fig.~\ref{fig:qnodelink} depicts an example of a small quantum repeater network consisting of QNodes connected by quantum channels.
QNodes are physically connected by a quantum communication-capable channel, such as optical fiber.
Adjacent QNodes can create entanglement between their qubits. We assume that QNodes can classically communicate with any other QNodes via classical channels such as the Internet (not shown in Fig.~\ref{fig:qnodelink}).

\Figure[htb][width=240pt]{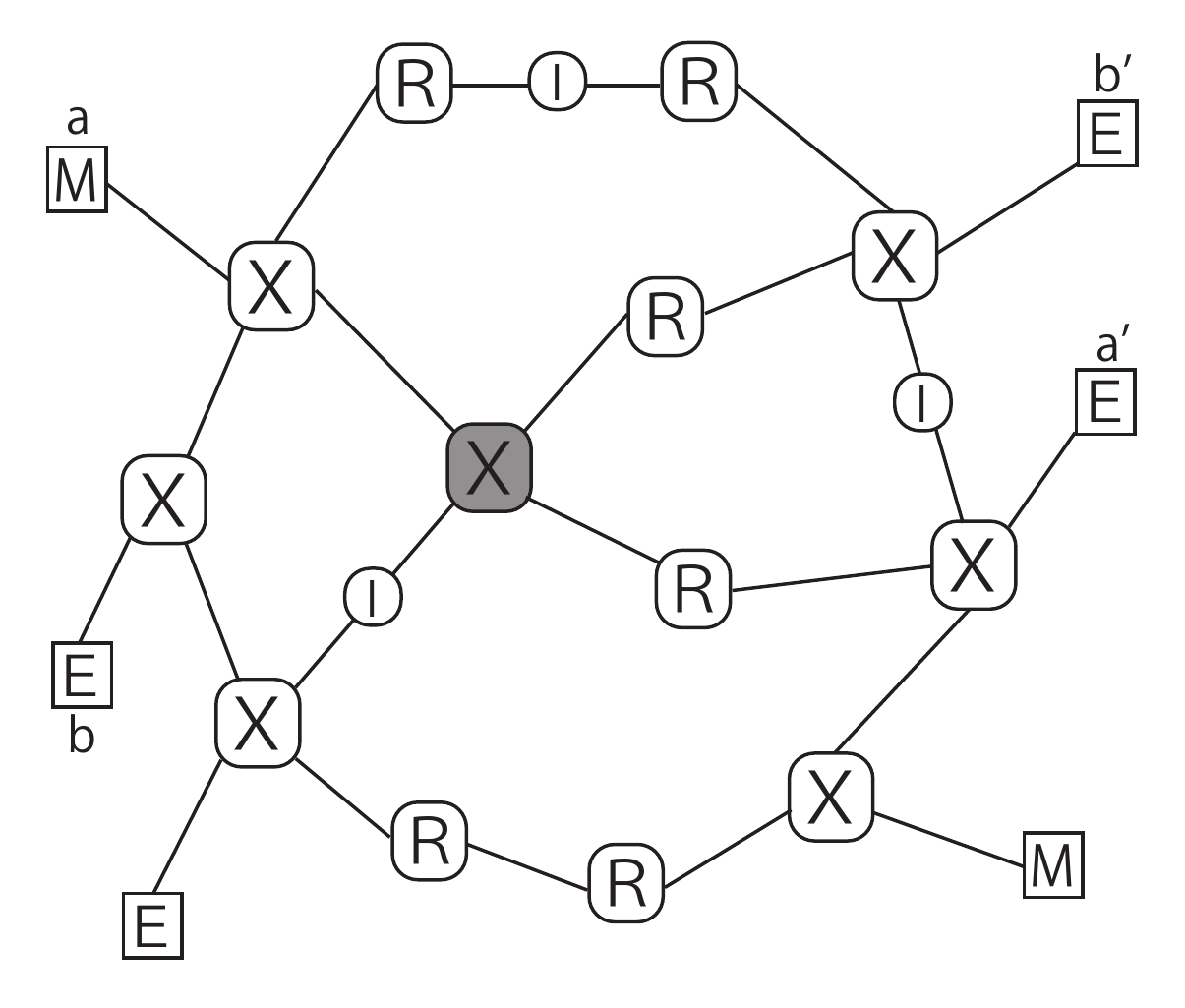}
{Schematic diagram of interconnected nodes. Capital letters indicate the first letter of the QNode type. Bell pairs can be shared directly between adjacent nodes. Classical communication is possible between all nodes. Distances will depend on technology, but most likely will be 10s of \si{km} over fiber. If the central (dark) XNode is lost due to hijacking or failure, the number of hops between a-a' (b-b') will increase, and rerouting will strain the bandwidth of the other links~\cite{satoh2018network}.
\label{fig:qnodelink}}

Each QNode has (classical) network addresses, such as IP addresses, for various inter-repeater classical information communication. 
For this we may use the current Internet, or we may need to build a new dedicated classical network that is strictly isolated and monitored, accessible only by authorized parties.
In any case, our minimum requirement is having an address unique among the set of reachable quantum repeater nodes.
We assume each
quantum node has a unique address. 
Since all quantum nodes require
both quantum and classical communication, a natural approach is to use
global IP addresses as an addressing scheme.  
This is also the most general, from the point of view of security analysis.
To simplify our discussion, we assume that all repeaters are connected to some kind of classical network that allows them to communicate with each other.

Topics that may require consideration in the context of a global-scale classical Internet, such as distributed denial of service (DDoS)-style attacks, are outside the scope of this paper.

\subsection{Link types and generations of quantum repeater}
\label{subsec:typeqn}
Muralidharan et al. defined three generations of quantum repeaters based on qubit transfer schemes and required technical level~\cite{muralidharan2016optimal}.
In first and second generation quantum repeater networks  each pair of directly connected QNodes needs to share Bell pairs as the first step of quantum communication.
A Bell pair is shared between non-adjacent QNodes by entanglement swapping operation at the relay QNodes. 
In the third generation quantum repeater networks, logical states encoded directly in a large number of photons are forwarded to non-adjacent nodes.

\subsubsection*{Entanglement swapping (ES) quantum repeater network} 
This type of network executes entanglement swapping to share Bell pairs between non-adjacent repeaters.
Some networks categorized in this type use non-encoded Bell pairs~\cite{briegel98:_quant_repeater, repeater2, Jiang30102007, RevModPhys.83.33} as first generation repeaters~\cite{muralidharan2016optimal} and some employ quantum error correcting codes~\cite{PhysRevLett.104.180503,VanMeter_2009,1367-2630-15-2-023012,knill96:concat-arxiv} as second generation quantum repeaters.
Repeaters transfer quantum information by teleportation using shared Bell pairs.

\begin{figure*}[htbp]
    \begin{tabular}{c}
    \begin{minipage}{.33\hsize}
        \centering
        \includegraphics[width=39mm]{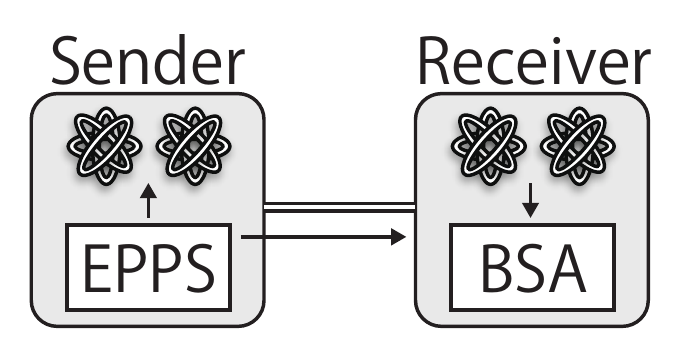}
        \subcaption{Memory to Memory link}
    \end{minipage}
    \begin{minipage}{.33\hsize}
        \centering
        \includegraphics[keepaspectratio, width=53mm, angle=0]{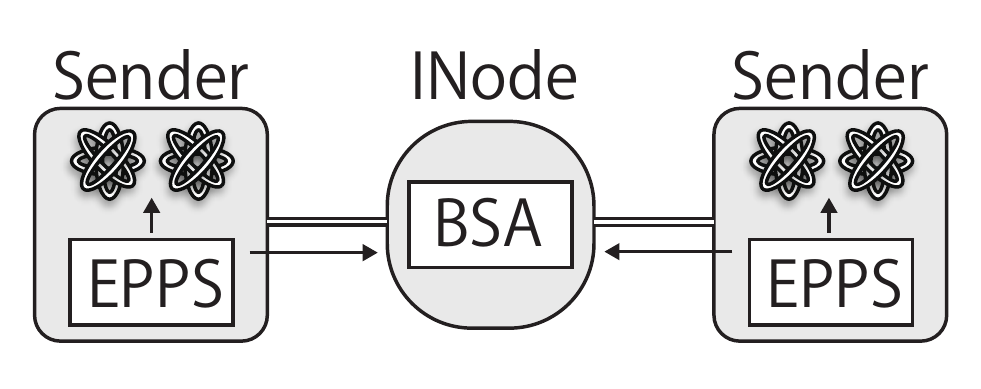}
        \subcaption{Memories and BSA link}
    \end{minipage}
    \begin{minipage}{.34\hsize}
        \centering
        \includegraphics[keepaspectratio, width=54mm, angle=0]{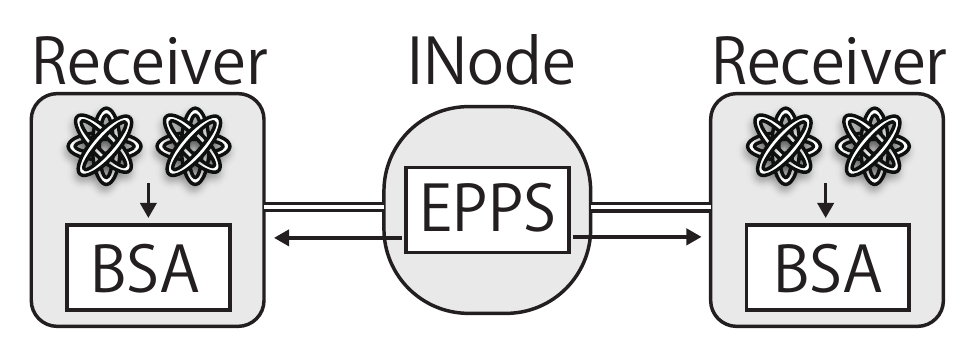}
        \subcaption{Memories and EPPS link}
    \end{minipage}
    \end{tabular}
    \caption{Model of ES-type QNode-to-QNode connections. Arrows denote the movement of one half of the Bell pair.}
\label{fig:qtq}
\end{figure*}
As shown in Fig.~\ref{fig:qtq}, we modeled ES-type QNode-to-QNode connections into three models, based on their schemes for creating Bell pairs~\cite{jones2016design}:
\subsubsection{Memory to Memory link}
This type consists of two QNodes.
One QNode receives a photon from a connected RNode (or ENode) and creates
Bell pairs using a BSA~(Bell states analyzer) entangler built into the node.

\subsubsection{Memories and BSA link}
This type consists of two QNodes and
one standalone BSA at the midpoint of the link.
The BSA receives a photon entangled with an interface qubit from each QNode, and creates
Bell pairs between the interface qubits by measuring the photons together, using the BSA entangler.

\subsubsection{Memories and EPPS link}
This type consists of two QNodes and one standalone EPPS
at the midpoint of the link. An EPPS creates photonic Bell pairs and transmits them to connected RNodes (or ENodes)~\cite{jones2016design}. 

\subsubsection*{Direct transfer (DT) quantum repeater network}
In this type, repeaters directly send encoded quantum states~\cite{PhysRevLett.112.250501} and are categorized as third generation quantum repeaters.
The individual links must have very high success probability of transferring physical qubits.

\subsection{Management of a quantum repeater network}
Topology, routing, multiplexing or allocation of network resources, and monitoring of links and the network itself are key work items for network nodes.  All of these functions must be designed and implemented assuming the presence of malicious actors.
The details of topology and routing are beyond the scope of this paper.
We focus on handling malicious actors.
The essential countermeasure for malicious actors is the same as the technology used for network monitoring, broadly centered around the concept known as \emph{quantum certification}~\cite{eisert2020quantum}.

\subsubsection*{Quantum certification}
Quantum certification aims to characterize the state created by a particular quantum device or process.
Approaches to quantum certification vary based on the amount of resources they consume, the amount of information they provide about the state and crucially the assumptions that are made about the devices being tested \cite{eisert2020quantum}.
The traditional example of a certification approach is quantum state tomography \cite{altepeter2005photonic,cramer2010eqs,james2001moq}.
Tomography provides a large information gain about the state and therefore can be used to make accurate estimates of the state fidelity.
This comes at the cost of requiring a tremendous amount of resources  as well as making strong assumptions about the quantum devices being tested.
For these reasons quantum state tomography may play a useful role only in the early stages of the Quantum Internet as shown in Tab.~\ref{table:qi_stage}.
As the technology improves, the demands on the security will increase.
In particular, it will not be possible to assume trusted devices as is the case in quantum state tomography.
Device-independent protocols \cite{liu2018device,reichardt2013classical,hajdusek2015device,hayashi2018self} only assume that the tested devices are governed by the laws of quantum mechanics and therefore guarantee greater security.
An important ingredient of such an approach is self-testing \cite{supic2020selftestingof} where violation of a Bell inequality is used as a guarantee of high fidelity of the distributed states as well as their \Conf \cite{PhysRevLett.23.880,horodecki2009quantum,koashi2004monogamy,terhal2004entanglement,pathumsoot2019modeling}.

Quantum certification requires the generation and consumption of many Bell pairs to determine the statistical characteristics of a quantum channel or path, and cannot be used to determine anything about any individual Bell pair.
An important requirement is that the selection of Bell pairs to be sacrificed for  certification must be random and secure;
if the eavesdropper can predict which pairs will be used, she can
remain undetected simply by choosing not to entangle or interfere with
those pairs~\cite{satoh2018network}.
Confidence in our assessment grows slowly as a full tomographic procedure converges incrementally by consuming substantial numbers of Bell pairs, so other approaches to state monitoring are under development; this remains an important research topic for robust, secure, efficient Quantum Internet operation~\cite{eisert2020quantum,oka16:qcit}.
Violation of a Bell inequality supporting quantum certification serves as the basis of one form of quantum key distribution~\cite{ekert1991qcb}.
Hence it is secure, given an authenticated classical channel and classical plane.
Considering the physical implementation, the qubits used for certification must be isolated from the external network in this type of verification scheme, as cracking of the Bell test through the photon detector has been demonstrated~\cite{Jogenforse1500793}.

\subsubsection*{Classical authentication}
Classical channel between QNodes performing quantum teleportation or quantum certification requires authentication.
If we perform quantum certification without authentication, the classical communication partner that shares measurement results and basis information may not be the original QNode, but the attacker who is stealing qubits.
In this case, we obviously cannot detect the attack by quantum certification.
And then we would teleport the quantum data to the attacker.
The \Conf of quantum data requires authenticated classical communication, and we assume this for the discussion in Sec.~\ref{sec:primitive} and~\ref{sec:pl}.

\subsection{Applications of a Quantum Internet}
\label{subsec:qiapp}
The ultimate purpose of the Quantum Internet, which consists of
distributed QNode connected with both quantum
channels and classical channels, is to create entanglement between two or more
terminal application qubits in two or more distant QNodes
chosen at the discretion of the application user.  
Based on the level of required functions, Wehner {\it et al.} classified the development of a Quantum Internet by stage and showed the applications provided at each stage~(Tab.~\ref{table:qi_stage}~\cite{Wehner18:eaam9288}).
This paper focuses on networks at stage 2 and above.

\begin{table}[htb]
    \centering
    \begin{tabularx}{\linewidth}{l|X}
        {\bf Stage of Quantum Internet} &  {\bf Examples of known applications} \\ \hline
        1.~Trusted repeater & QKD (no end-to-end security)\\
        2.~Prepare and measure & QKD, secure identification \\
        3.~Entanglement generation & Device independent protocols \\
        4.~Quantum memory & Blind quantum computation, simple leader election and agreement protocols \\
        5.~Few qubit fault tolerant & Clock synchronization, Distributed quantum computation\\
        6.~Quantum computing & Leader election, fast byzantine agreement\\
    \end{tabularx}
    \caption{Stages in the development of the Quantum Internet~\cite{Wehner18:eaam9288}. As the stage progresses, more advanced hardware is required to deliver richer functions.}
    \label{table:qi_stage}
\end{table}

Application layer protocols may have end-to-end verification mechanisms, such as QKD and more general device-independent protocols (e.g. Above classical authentication via another provider service).
On the other hand, an attack on the routing layer is a sufficient threat even if the application layer adopts secure protocols in classical networks~\cite{10.1145/3429775}.
From this point of view, we believe that not all security to \Conf should depend on the application layer, and  we mainly discuss attacks outside the application layer.

\section{Hardware model of the Quantum Internet}\label{sec:model}
\begin{figure*}[htbp]
    \begin{tabular}{c}
    \begin{minipage}[t]{.25\hsize}
        \includegraphics[keepaspectratio, width=42mm, angle=0]{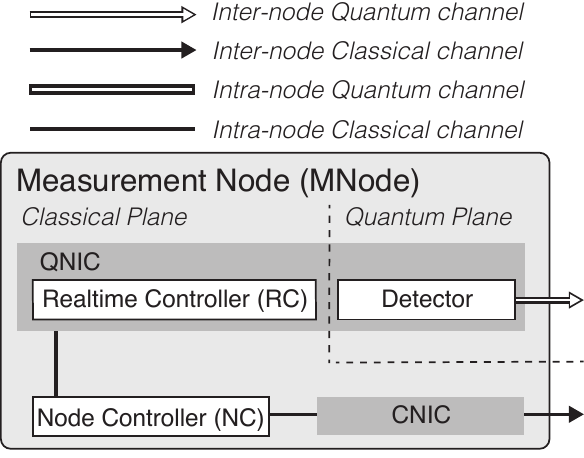}
        \subcaption{}
        \label{fig:qmnode}
    \end{minipage}
    \begin{minipage}[t]{.24\hsize}
        \includegraphics[keepaspectratio, width=40mm, angle=0]{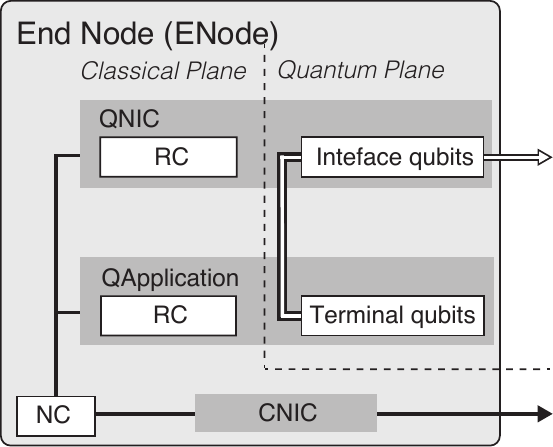}
        \subcaption{}
        \label{fig:qenode}
    \end{minipage}
    \begin{minipage}[t]{.21\hsize}
        \centering
        \includegraphics[keepaspectratio, width=34mm, angle=0]{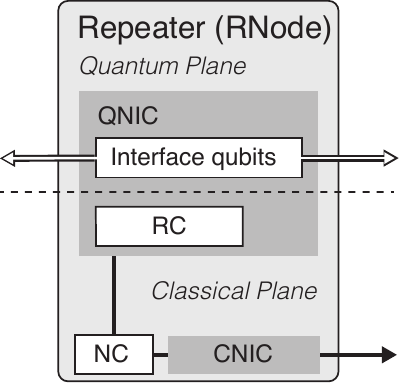}
        \subcaption{}
        \label{fig:qrnode}
    \end{minipage}
    \begin{minipage}[t]{.25\hsize}
        \centering
        \includegraphics[keepaspectratio, width=46mm, angle=0]{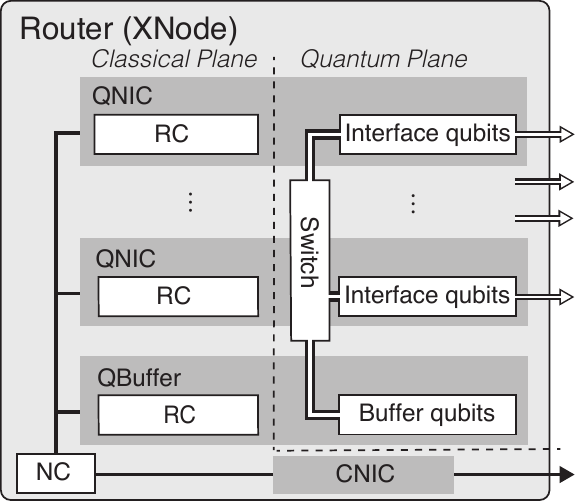}
        \subcaption{}
        \label{fig:qxnode}
    \end{minipage}
    \end{tabular}
    \caption{Model of QNodes. (a) MNode must be able to  measure incoming qubits in any basis. Like other QNodes, we can separate components into Quantum plane and Classical plane. (b) ENode can perform universal computations on terminal qubits.  (c) An RNode connects two non-adjacent QNodes. (d) XNode connects to several QNodes and is responsible for communicating on various routes.}
\label{fig:qnodes}
\end{figure*}

In this section, we describe our models of QNodes and
those elements. A QNode will have different modules depending on its role shown in Sec.~\ref{sec:nodetype}.
We draw the internal structure of each QNode type in Fig.~\ref{fig:qnodes} (see also Sec.~\ref{subsec:Q2Q} for INode).
We classify these components into two planes, classical and quantum, and describe details as follows.

\subsection{Quantum plane components of a QNode}
The networks corresponding to stages 1 to 4 in Tab.~\ref{table:qi_stage} are functionally constrained, and we call them NISQI, or Noisy Intermediate Scale Quantum Internet.  In this era, qubit memories will be noisy and limited in number.  In fulfilling the network's responsibilities, they will have several roles.  Ideally, these roles are performed by qubits with different physical placement in the system, but with limited resources qubits may be assigned more than one role.  With the primary focus of this paper being on physical vulnerability to attacks, for security analysis purposes, qubits should be treated as the most vulnerable role to which they belong.  For example, buffer qubits (described below) are defined to be isolated from the inter-node optical channel; if software assigns the role of buffer to an interface qubit, it still must be treated as an interface qubit.

\subsubsection*{Quantum Network Interface Card (QNIC)}
A QNIC is a quantum network's equivalent of a classical NIC (Network Interface Card).
Depending on the physical implementation, it may consist of
transmitters, receivers or detectors, and qubits (Interface qubits) used to create entanglement with a remote QNIC's qubits.
A QNIC has both internal and external interfaces.
An internal interface consists of both control and quantum connections
to other elements in the QNode.
An external interface is a quantum channel, combined with basic, hard real-time
classical signaling for framing and sequencing.
A QNIC will be connected to a counterpart QNIC with a physical link such as a fiber.

A hard real-time controller (RC) in a QNIC also handles all real-time operation, such as automatic creation of on-physical-link entanglement~\cite{reilly2015engineering}.
In Fig.~\ref{fig:qnodes}, we control each classified Qubits with a different RC, but a single integrated RC could be adopted.

The node controller can direct the QNIC to operate on interface qubits.

\subsubsection*{Quantum Application Platform (QApplication)}
A QApplication controls terminal qubits, intended for quantum applications (see Sec.~\ref{subsec:qiapp}).
These qubits can be entangled or swapped with interface qubits, and hence can be entangled with remote terminal qubits.

\subsubsection*{Quantum Buffer (QBuffer)}
A QBuffer is a pool of qubits (Buffer qubits) used to prevent QNIC congestion.
A QBuffer may be optional depending on workload and
hardware design, but our analysis assumes it is present.

\subsubsection*{Interface qubits}
Each QNIC other than MNode has multiple interface qubits
which are exposed to inter-node channels and can create entanglement with remote interface qubits.
Interface qubits in QNICs are used only temporarily.
Once the entanglement which the interface qubits hold is transferred to buffer qubits or terminal qubits, interface qubits play no further role in that operation.

\subsubsection*{Buffer qubits}
Buffer qubits in a QBuffer (usually in a repeater or router) can hold entangled states but are not optically connected to the outside, freeing interface qubits for reuse and are more secure than interface qubits.
Buffer qubits are used for temporal buffering.
They may have different physical characteristics than
interface qubits and can interact with either interface qubits or terminal qubits.

\subsubsection*{Terminal qubits}
Terminal qubits are similar to buffer qubits, but are located at ENodes.  Terminal qubits are under the direct control of a QApplication.  They are assumed to be optically isolated from any inter-node quantum channel.

\subsubsection*{Detector}
A detector is a device for measuring received photonic qubits.  Here, detectors are assumed to be non-photon number counting detectors; they detect the presence or absence of light.  Measuring a photonic qubit in a specific basis is different depending on the photonic qubit representation (e.g., time bin or polarization), and may require the addition of linear optical elements such as polarizing beamsplitters.  Detectors are often built into larger subsystems such as Bell state analyzers. Detectors are connected through quantum channels to buffer qubits, terminal qubits or interface qubits, either via an optical switch or directly.
Detection is a conversion from quantum information to classical information.

\subsubsection*{Intra-node quantum channel}
Intra-node quantum channels provide interconnection between quantum elements in a QNode such as terminal qubits, buffer qubits and interface qubits.
Intra-node quantum channels are not exposed to the outside of the node.

\subsubsection*{Inter-node quantum channel}
By using an inter-node quantum channel, node-to-node single hop entanglement is created between interface qubits.

\subsubsection*{Linear optical components}
Passive linear optical components, such as ordinary or polarizing beamsplitters or wave plates, alter photonic qubit states in simple ways or are combined with other components to create subsystems such as Bell state analyzers.

\subsubsection*{Optical switch}
An optical switch (e.g., a nanomechanical crossbar~\cite{kim03:_1100_port_mems}) changes the optical connections between quantum plane elements.  Although such switches can be used in the long-distance optical paths~\cite{elliott02:_building_quant_network}, here we focus on use inside a node with multiple QNICs to achieve non-blocking photon routing. A node with one or two QNICs would not need a node-internal Switch (Fig.~\ref{fig:qnodes}).

\subsection{Classical plane components of a QNode}
\subsubsection*{Classical Network Interface Card (CNIC)}
A CNIC is a standard classical network interface that can be connected to the classical Internet. We assume this to be an interface such as Ethernet.
The CNIC provides inter-QNodes communications,
generally carrying soft real time information necessary for
interpreting quantum information and determining future operations.

\subsubsection*{Real-time Controller} A real-time controller controls the qubits in each unit, meeting the hard real-time constraints
for maintaining quantum states and performing operations on qubits either individually or collectively.  In our current
model, three types of real-time controller are shown: the QNIC real-time controller, the Buffer real-time controller and
the Application real-time controller.

\subsubsection*{Node Controller} The Node Controller communicates with other QNodes and controls QNIC, QBuffer and QApplication to
achieve its goal: for a repeater node, to create entanglement between a local interface qubit and a remote
interface qubit;  for an ENode, to create entanglement between a local terminal qubit and a remote qubit
via its interface qubit, to run an application.

\subsubsection* {Intra-node classical channels}
Here we stick to intra-node classical channels as a subsystem of intra-node qubit transmission functionality, such as transferring meta-information, Pauli frames (which define the polarity of a qubit relative to a reference signal) or acknowledgments of quantum channels.
Intra-node classical channels are not exposed to the outside of the node.

\subsubsection*{Inter-node classical channels}
Inter-node classical channels are used to coordinate with other nodes.

\subsubsection*{Other classical computing elements}
Since a QNode consists of hybrid classical computing elements and
quantum elements, it also may have various classical computing elements
such as clock, memory, processor, and chassis including expansion buses or
backplanes.

\subsection{Elements of QNode to QNode connection and external resource}
\label{subsec:Q2Q}
\subsubsection*{Entangled Photon Pair Source (EPPS)}
An EPPS may be connected to inter-node quantum channels in intermediate nodes, or may be used along an intra-node quantum channel in other nodes.
An EPPS is a simple component which continuously creates and sends entangled photons to two connected components.  The canonical example of an EPPS is \emph{symmetric parametric down conversion} (SPDC)~\cite{PhysRevLett.75.4337}.

\subsubsection*{Bell state analyzer (BSA)}
A BSA is a subsystem built from detectors and linear optical components that measures two photons in the Bell basis. If the incoming photons are entangled with stationary memories or other photons, the measurement effects entanglement swapping.  A BSA may be deployed as a standalone INode or incorporated into a QNode.

As mentioned in Sec.~\ref{sec:nodetype}, when we use BSA and EPPS as a standalone nodes (INodes), they also have inter-node classical channels.
BSA implementations using linear optics suffer from the shortcoming that only two of the four Bell states can be distinguished with certainty.
This limits the success probability of such BSA to $50\%$.
Also, the canonical BSA cannot choose the outcome of the measurement; the two photons are projected onto one of the Bell states at random.

\subsubsection*{Quantum external connectivity}
All QNICs are connected to other adjacent QNICs (or QNodes) via optical link.
Such a link is point-to-point system for creating entanglement~(see Sec.~\ref{subsec:typeqn}).

\subsubsection*{Classical external connectivity}
Through a CNIC, QNodes are connected to private network or the classical Internet.
All QNodes may communicate with each other via this external classical connectivity.
\section{Attacks without Control of Quantum Nodes}\label{sec:primitive}

In this section, we describe the primitive attacks on each element of the Quantum Internet.
Quantum network devices can be divided into the quantum plane and the classical plane.
The quantum plane holds qubits and quantum channels. 
The classical plane holds quantum application software, qubit controllers, classical channels, and operating systems.
Primitive attacks come externally.
(More complex attacks may be caused internally by the hijacked components, including physical security violation.)

This section summarizes how primitive attacks affect the security notions of CIA (\Conf, \Integ, \Avail).
The CIA of the nodes and links introduced in Sec.~\ref{sec:model} can be considered by combining elemental discussion of primitive attacks in this Section.
Due to the no-cloning theorem, \Conf, \Integ and \Avail are closely linked.
For example, by stealing quantum data from an interface qubit, another value can be injected in the qubit. 
In this case, an attack on \Conf attacks \Integ, too.
The focus of this paper is categorizing attacks, hence such relationships are beyond the scope.

We note that the attacks covered in this section are based on physical manipulation of or physical proximity to each element.
\subsection{Attacks on the Quantum Plane}
First, we will discuss the target devices on the quantum plane and the details of each attacking scheme shown in Table~\ref{table:qp_attack}.
\begin{table*}[htb]
    \centering
    \begin{tabularx}{\linewidth}{l|X|X|l}
        {\bf Attacker's action} & {\bf Attacker's resources} & {\bf Components under attack} & {\bf Compromised elements} \\ \hline
        Eavesdropping & Access to inter-node channel & In-flight optical states & Confidentiality, Availability \\
        Optical probe & Insertion \& detection of optical pulses & Interface qubit, Intra-node Quantum channel, Detector & Confidentiality\\
        Fault injection & Insertion of optical pulses & Interface qubit, Quantum channel (Intra-node, Inter-node), Detector  & Integrity, Availability\\
        Out-of-system, standoff attacks & Physical access near nodes & All Qubits, Quantum channel (Intra-node, Inter-node), Switch, Detector & Integrity, Availability\\
        Malicious entanglement & Ersatz node & Interface qubits & Confidentiality, Integrity, Availability\\
        Destruction or vandalism & Physical access to node equipment & Interface qubit, Intra-node quantum channel, Switch, Detector & Availability\\
        Theft of hardware & Physical access to node equipment, node operational expertise & All & Confidentiality, Integrity, Availability
    \end{tabularx}
    \caption{Attacks on the quantum plane, ordered roughly according to the severity or intrusiveness of the attack. }
    \label{table:qp_attack}
\end{table*}

\subsubsection{Eavesdropping (quantum channel)}
  {\bf Eavesdropping} on photons flying in {\bf inter-node quantum channel} affects \Conf if valuable quantum data is encoded onto photons. Transferring half Bell pairs and executing quantum teleportation afterwards would protect valuable quantum data from eavesdropping as long as the operation is authenticated properly.

  {\bf Eavesdropping} also works as an attack on \Avail, because it breaks the quantum states.
  This Denial-of-Service attack is one of the most obvious weaknesses of quantum networks if robustness is an important design goal.
  Since an inter-node quantum channel is just a fiber or such cable between QNodes, it is relatively easy to get access to these channels.

\subsubsection{Optical probes} 
  {\bf Optical probing} of detector settings has been demonstrated.
  Among the many attacks on QKD implementations developed in Makarov's lab, Jain \emph{et al.} described an eavesdropper that can probe the chosen measurement basis used in a BB84 quantum key distribution (QKD) system~\cite{bb84} by sending a bright pulse of light from the quantum channel into the interface and analyzing the back-reflected pulses~\cite{jain2014risk},
  a classical attack on the hardware used for the quantum states.
  The attack could be executed directly by cutting fibers and inserting hardware, or by tapping the fiber e.g. via evanescent coupling.
 
  Though entanglement-based QKD protocols do not have this weakness because state certification is performed end-to-end, a similar attack in which some optical detectors are saturated could be used in a man-in-the-middle attack.
 {\bf Optical probes} damage the \Conf of quantum states and the following devices will also be attacked:

\subsubsection*{Case: Detector with BSAs}
For XNodes, those (backplane) detectors are located behind the node-internal optical switch and are not exposed.
Hence {\bf Optical probes} do not affect \Conf.

For intermediate nodes, these detectors are optically connected to the inter-node channel directly and are exposed.
Then, {\bf Optical probes} may be used to determine hardware settings, or may be used to control what the classical hardware sees.
  Nevertheless, \Conf will not be affected, as quantum state leakage is not expected to occur.

\subsubsection*{Case: inter-node quantum channel}
{\bf Blinding of detectors} as a kind of {\bf Optical probe} would cause the leakage of quantum state, while the detector saturation attack described above could be used to control what the classical hardware sees.
More analysis of this impact on QNode operation is necessary.
Since an inter-node quantum channel is physically just an optical fiber or similar media between QNodes, it is relatively easy to get access to these channels.
  
{\bf Optical probes} are not executable on the {\bf intra-node quantum channel} without directly modifying the hardware.

\subsubsection{Fault injection}
  {\bf Fault injections} involve inserting unauthorized and unexpected optical pulses into the inter-node quantum channel, hoping to reach the detector~\cite{Benoit2005,Lemke2005}.
  For some qubit representations, it is also known that altering the temperature of fibers affects the quantum states of photons passing through by slightly altering the effective path length.
  
Depending on the design of the QNodes, {\bf fault injections} may disturb the state of {\bf interface qubits}, {\bf inter-node quantum channel}, and {\bf detector} (if exposed), affecting \Avail and \Integ.
    This attack is also not executable on the intra-node quantum channel without directly modifying the hardware.
    
\subsubsection{Out-of-system attacks}
  {\bf Out-of-system attacks} such as direct irradiation of a device with RF noise could damage the quantum data and leave garbage.
  Such attacks may act as any quantum operation, including non-computational operations such as leakage from the computational basis.
  For this kind of attack, an attacker may not even need access to the target device itself, as radio waves can blanket an area from a modest distance.
  Even with good RF shielding, interference effects as weak as subway power and control systems a kilometer away are known to affect some systems.
  Other attacks, such as on the cooling or other control systems, may be harder to carry out remotely.  
  
  These attacks also prevent designed qubit operations, which affects \Integ and \Avail of interface qubits, buffer qubits and terminal qubits.
  {\bf Out-of-system attacks} would alter the state of photons during transfer and correct signals become unrecognizable due to injected noise. Therefore, intra-node quantum channel, switches, inter-node quantum channel, and detector would also be targets of such attack.
    
\subsubsection{Malicious entanglement}
{\bf Malicious entanglement} attempts to steal information via teleportation by replacing a legitimate qubit with a qubit created by and entangled with other qubits controlled by the attacker, or by entangling additional qubits with otherwise legitimate qubits under the nominal control of legitimate nodes.
In the worst case, this malicious entanglement can result in the attacker being the receiving party in a teleportation operation; this theft of a qubit causes the loss of \Conf.
Malicious entanglement also disturbs  \Integ because it affects any other qubits with which it is entangled, and affects \Avail as a result of the no-cloning theorem~\cite{no_cloning}.

As creating malicious entanglement requires that the attacker have control of some quantum memory, this involves either compromise of an otherwise licit network node, or connection of an illicit, ersatz network node.  Compromised control of licit nodes will be addressed further in Sec.~\ref{sec:pl}.

The targeted qubit operations and the affected situations are described below:
\subsubsection*{Case: Interface qubit}
Since optical fibers for inter-node quantum channel are attached to {\bf interface qubits} directly (see below), interface qubits are most exposed to external risks.
{\bf Malicious entanglement} may affect interface qubits via the inter-node quantum channel, by receiving maliciously entangled qubits or by having half Bell pairs sent from this interface entangled with malicious qubits by eavesdroppers' operations.
Malicious entanglement would result in the theft of valuable quantum data if quantum teleportation is executed without awareness of the attack.

The essential countermeasure for this vulnerability is quantum state certification.
Assuming interface qubits are used to temporarily hold half Bell pairs (completely generic states with no secret information) before teleporting valuable quantum data, quantum certification \emph{randomly} selects some of the Bell pairs to measure to determine if an eavesdropper has entangled her qubits with ours, as described in Sec.~\ref{sec:qrintro}.

  We may consider the scenario where the eavesdropper avoids detection by sheer luck and interacts with qubits which were not picked for the certification procedure.
  In this case, the \Conf of quantum data is not necessarily compromised.
  If the classical channel retains its \Conf, so does the quantum data.
  Classical data must be sent to the recipient in order to successfully complete teleportation of quantum data.
  Without this classical data the eavesdropper ends up in possession of useless maximally mixed states.

\subsubsection*{Case: Buffer qubit and Terminal qubit}
Since buffer qubits and terminal qubits themselves do not have any external connectivity, they are not exposed to risks from inter-node quantum channels directly.
However, since it is assumed that the quantum application can execute any quantum operation on the terminal qubits, a compromise of the application software is a compromise of the terminal qubits.

Data in a classical memory buffer can be assumed to be ``safe'', untouchable from the outside world provided the buffer cannot be reached by DMA hardware that can be activated from outside and the host OS has not been compromised.  
Our quantum data are similarly safe from fault injection once stored in terminal qubits.
Even if an eavesdropper has entangled a qubit of hers with our qubit before it reaches this buffer, she gains no access to information she did not already have at the time she entangled her qubit with ours.
Quantum certification while working with a stream of Bell pairs is needed here as well as on interface qubits.

\subsubsection*{Case: NISQI qubit}
NISQI qubits will have all of the weaknesses of other qubits due to their many roles.
This assumption is likely to be true for the other attack methods discussed below, and attackers will have many avenues to exploit.

\vspace{0.1in}
{\bf Malicious entanglement} may be inserted via interface qubits, then transferred to terminal qubits, or other buffer qubits. 
If the insertion is successful, malicious entanglement is threats against \Conf, \Integ and \Avail of buffer qubits, terminal qubits as well as interface qubits.

\subsubsection{Destruction or vandalism of hardware}
{\bf Destruction of or damage to hardware} and other typical classic attacks prevent the designed operation of the qubits, if the attacker has access to the target device.
These attacks pose a threat to the \Avail of each QNode device: qubits (interface, buffer, terminal), intra-node quantum channel, optical switch, and detectors with BSA.

\subsubsection{Theft of hardware}
Theft of quantum equipment naturally affects \Avail of the system and \Integ of any data active at the time of theft, but \Conf is not affected, because quantum systems do not yet support non-volatile memory.  The state of any memory would be destroyed by removal of power to the system as part of a theft.

\subsection{Attacks on the Classical Plane}
Next, we will discuss the target devices on classical plane and details of each attacking scheme shown in Table~\ref{table:cp_attack}.  
The attacks here concentrate on reading, modifying or destroying classical messages and signals in the channel, but extend to more physical attacks as well.
Some of these attacks aim to take control of a quantum node via classical vulnerabilities; cases in which such hijacking has succeeded are discussed in the next section.

\begin{table*}[btp]
    \centering
    \begin{tabularx}{\linewidth}{l|X|X|l}
        {\bf Attacker's action} & {\bf Attacker's resources} & {\bf Components under attack} & {\bf Compromised elements} \\ \hline
        Eavesdropping & Access to inter-node channel & Data-relevant messages & Confidentiality\\
        Message disruption & Access to inter-node channel & RC, Detector with BSA, EPPS & Availability\\
        Man-in-the-middle attack& Intercept \& resend on inter-node channel & All connection setup, action messages, network management tomography, etc. & Integrity\\
        Message rerouting & Control of a classical router & Switch, Detector with BSA & Confidentiality, Integrity, Availability\\
        DoS attack & Ability to overtax network resources & Intra-node classical channel & Availability\\
        Destruction or vandalism & Physical access to classical controller & CNIC, Controller resources, Electric power & Availability\\
        Theft of hardware & Physical access to node equipment, node operational expertise & All & Confidentiality, Integrity, Availability\\
    \end{tabularx}
    \caption{Attacks on the classical plane ordered roughly according to their invasiveness or severity.  
    }    \label{table:cp_attack}
\end{table*}
  
\subsubsection{Eavesdropping (classical channel)}
All classical attacks aimed at {\bf inter-node classical channels} may be possible.
We will show examples of specific attacks on classical channels in quantum network systems, for each compromised element.

Classical privacy threats, such as {\bf eavesdropping} and {\bf tracking} user behavior, would progressively reduce the inherent \Conf of quantum information.

\subsubsection{Message disruption}
The disruption of classical communications will destroy \Avail.
There are concrete cases such as {\bf Attacks to clock synchronization} against {\bf real-time controller}, {\bf EPPS}, and {\bf detectors with BSAs} and {\bf Attacks to measurement information} against {\bf real-time controller}.

\subsubsection{Man-in-the-middle attack}
  {\bf Man-in-the-middle attack} against {\bf inter-node classical channels} can disrupt the generation of Bell pairs, in a variety of ways, including overwriting Pauli frames.
  It would be difficult to determine from the quantum state disturbance whether this attack, which violates \Integ, is an attack on classical or quantum communication.
  
\subsubsection{Message rerouting}
  {\bf Message rerouting}, disobeying routing information, would forward qubits to eavesdroppers.
  This attack is achieved by {\bf falsified routing information} as well as hijacked controller of {\bf optical switches.}

  {\bf Message rerouting} would work as {\bf malicious entanglement} and may result in inserting incorrect quantum states, thus affecting \Integ and \Avail.
  In a particular configuration of the network, using bounded entanglement state (or low distillation state) we can resist such an attack~\cite{PhysRevResearch.2.043022}.

\subsubsection{Denial-of-service (DoS) attack}
When using public classical channels such as the Internet,  {\bf Denial-of-Service attacks} would disrupt coordination messages such as ACK messages for photons, affecting \Avail.

\subsubsection{Destruction or vandalism of hardware}
These attacks are well studied and explained by e.g. Weingart~\cite{weingart2000psd}.
In our networks, the following classical computing elements responsible for \Avail can be attacked:
\begin{itemize}
\item CNIC
\item Controller resources (such as clock, memory, processor)
\item Chassis providing electric power.
\end{itemize}

\subsubsection{Theft of hardware}
Theft of the classical equipment controlling quantum systems naturally affects \Avail of the system and \Integ of any data active at the time of theft.  Here, we distinguish theft from simple destruction (above) by defining theft to assume the attacker has the ability to operate the node. Unlike the quantum plane devices, classical components often contain non-volatile storage holding data important to security.  Encryption keys used for authentication or for keying communication sessions are of particular concern. Thus, while an immediate attack on the \Conf of in-progress communications is not possible here, looking forward, the attacker gains all of the capabilities ascribed to malicious nodes in the next section.

\section{Attacks Using Malicious QNode(s)}\label{sec:pl}
The game changes substantially if a malicious party gains control of one or more QNodes. In this section, we investigate what can be done by an attacker(s) who have successfully hijacked full control of one or more QNodes.
The attack vectors and protocol stack considered in this section are based mainly on architectural principles outlined in~\cite{satoh2018network} and~\cite{matsuo2019quantum}.
It is worth noting that the types of QNodes, along with their vulnerabilities, discussed in this section are expected to form the basic building blocks of the Quantum Internet.
Early realizations of quantum networks will not be implemented exactly as outlined here. For example, some node designs will have no physical distinction among types of qubits, but instead will only assign differences in roles via software. Nevertheless, even as different hardware designs evolve, the fundamental roles described here will form the architectural foundation. Those differences in implementation may result in some devices being inherently immune to certain attacks. This is a desirable feature, but one that should be recognized explicitly rather than going unstated, in order to ensure that all subsequent generations maintain such guarantees.

The Quantum Internet will continuously use a certain fraction of performance for cross-validation to detect anomalies such as equipment failures or hijacking~\cite{satoh2018network}.
\begin{itemize}
    \item Each QNode regularly and continuously verifies neighboring nodes and their assigned communications.
    \item The network performs these verifications at irregular intervals during normal communication and eventually detects the presence of an attacker.
\end{itemize}
Under these circumstances, what can an attacker do with the hijacked node before her presence is detected?

First, we discuss the case where an attacker successfully hijacks one or more QNodes and summarize attacks using these nodes in Table~\ref{table:qn_attack}.
\begin{table*}[htb]
    \centering
    \begin{tabularx}{\linewidth}{l|l|l|X}
        {\bf Attacker's action} & {\bf Attacker controls} & {\bf Components under attack} & {\bf Compromised elements} \\ \hline
        False failure report & MNode & Network reliability & Availability\\
        QDoS attack& ENode and MNode & Network resources & Integrity, Availability\\
        Malicious application & ENode & Terminal qubit & Confidentiality, Integrity, Availability\\
        Dishonest quantum computation &   ENode & Performed quantum computation & Integrity\\
        Link down attack & RNode & Network resources & Availability\\
        Link connection attack & RNode & Network resources & Integrity, Availability\\
        Man-in-the-middle attack & RNode & QKD operation & Confidentiality\\
        Switching disruption & XNode & Network reliability & Integrity, Availability\\
        Framing innocent repeaters & XNode & Local area network & Integrity, Availability\\
        Path black hole & Classical path management & Path setup information & Availability\\
        QDDoS attack& Multiple ENodes & Service providing QNode & Integrity, Availability\\
        Framing using multiple hijacked QNodes & Multiple QNodes & Wide area network &  Integrity, Availability\\
    \end{tabularx}
    \caption{Attacks using malicious QNode(s). }
    \label{table:qn_attack}
\end{table*}

\subsubsection*{Hijacked controller}
Since {\bf interface}, {\bf buffer}, and {\bf terminal qubits} are controlled by classical controllers, they can be hijacked by carelessness on the part of the controller (software or hardware).
An attacker will be able to arbitrarily control these qubits depending on the operations implemented, and will cause the following damage:
Loss of \Conf due to disclosure of quantum data via arbitrary sessions.
  Disruption of \Integ by changing values using manipulation of qubits (flipping qubits, measuring values, initialization, etc.).
  Loss of \Avail of any session due to improper execution or interruption of instructions.
  
  An attacker would use the hijacked {\bf intra-node classical channel} to cause the following threats:
Leakage of most meta-information, such as session information and classical channel usage~(\Conf).
Anomalies in the quantum state by tampering with Pauli frames and other management information~(\Integ).
Interference with specific sessions due to forged channel failures (\Avail).

 An attacker also could change the quantum state in the {\bf inter-node classical channel} by changing management information using a hijacked controller.

  {\bf Falsified measurement outcome} by hijacked {\bf detector with BSA} alters residual quantum states, affecting \Integ.
 {\bf Changing the input light power} of {\bf EPPS} would affect its \Avail.
  In addition to these active attacks, the {\bf hijacked controller} might also selectively interfere with the session by sabotage.

\subsection{Attacks by a malicious end user}
What can a malicious terminal node, an MNode or an ENode, do to attack the network?

\subsubsection{False failure report}
An attack by a MNnode would be nothing more than a false report claiming a network failure for reduction of \Avail.
If an attacker (falsely) complains that an application such as BQC will not work, the network will have to initiate careful self-verification. The network system can then conclude that either the claim is falsified or the MNode has failed.

\subsubsection{QDoS~(Quantum Denial-of-Service) attack}
A DoS~(Denial-of-Service attack) attack is a typical \Avail attack on the classical Internet. Attacks on network resources, such as network bandwidth and server computing power, will also work against the Quantum Internet.
Since the quantum state cannot be replicated~\cite{no_cloning}, the damage will be more serious than the classical DoS. In particular, we should be wary of direct qubit transmissions on DT repeaters (see Sec.~\ref{subsec:typeqn}), if the data qubits themselves are sent via the Quantum Internet. (If a half-Bell pair is sent followed by quantum teleportation, it does not matter.)
Examples of QDoS attacks include the following:
\begin{itemize}
    \item A malicious user requests that an oversized key be generated via quantum key distribution, overtaxing the quantum link bandwidth.
    \item A malicious user requests a massive number of calculations to a cloud quantum computer. Primarily an application server problem, this also affects network resource consumption and operations.
\end{itemize}
In order to counter these attacks, defenses in the classical plane, such as application behavior monitoring and request filtering, are important.
ENode and MNode are capable of these attacks.

Given physical access to an ENode, an attacker can easily entangle additional qubits into a state. This additional entanglement generally destroys the usefulness of a quantum state, acting as a QDoS attack. This attack is similar to attempts to eavesdrop on quantum key distribution. More formally, the intended qubits are subsystem of the larger entangled state, and hence are left in a mixed state.

\subsubsection{Malicious application}
\label{subsec:malapp}
{\bf Terminal qubits} are controlled by a quantum application controller.
Therefore, a compromise on the application side of the hardware affects terminal qubits.
Since the application controller is not a networking functionality, the detail of this loss of control is beyond scope of this paper.
However, such threats from applications need careful discussion.
{\bf Malicious application} may disclose the quantum data of  any session, disrupting the \Conf.
  Since fundamentally other components are not made to detect abnormalities of quantum applications, malicious applications are hard to detect.

\subsubsection{Dishonest quantum computation}
An ENode can carry out more sophisticated attacks to reduce \Integ.
When we perform distributed quantum computation, it is risky to commit part of the calculation to other ENodes.
An attacker could send incorrect quantum information via teleportation or perform wrong calculations.
It is challenging to ensure the redundancy of quantum information by the no-cloning theorem.
We should adopt a method that allows reliable quantum calculations even with a small number of fraudsters~\cite{PhysRevA.96.012303}.

\subsection{Attacks by a malicious repeater node}
An RNode can perform all three types of attacks that ENode~(MNode) can.
Fundamentally, an RNode participates in control of the two links to which it is connected and of the connections that pass through it.  Attacks performed using a hijacked RNode, then, affect one of these two subjects.

\subsubsection{Link attacks}
Obviously, a malicious RNode can simply mis-operate a link or mis-report the link status, bringing the link down altogether.  This results in connections routing around the node and link, resulting in network inefficiency.

More clever, insidious attacks may, for example, have the hijacked RNode perform the Bell test to check the link status in good faith, but not correctly perform the quantum operations required for inter-node communication.  This bad faith attack will cause the network to suspect that another node has failed.  This is a limited form of framing, discussed below.
To prevent more severe performance degradation, the network needs to perform a Bell test undetected by the attacker~\cite{satoh2018network}.

\subsubsection{Man-in-the-middle attack}
This attack technique, in which an attacker creates a fake private connection while eavesdropping and tampering, is popular in the classical Internet. The end nodes believe they have a single, private connection, but in reality each is communicating with the node in the middle.
If authentication is faulty, the attacker can also execute this attack in the Quantum Internet by using a hijacked RNode.
For example, we assume a ES type network for QKD operation using BBM92 (Fig.~\ref{fig:MITMA}) ~\cite{Bennett1992}.
\Figure[htb][width=240pt]{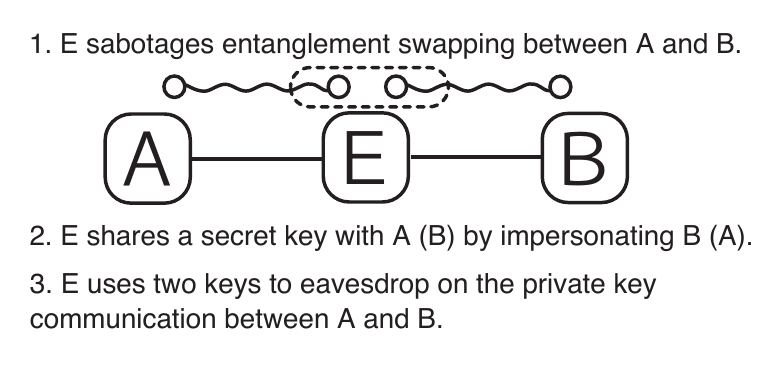}
{Man-in-the-middle attack on the Quantum Internet.
  \label{fig:MITMA}}
$A$ and $B$ share a Bell pair for quantum communication, so they ask $E$ for entanglement swapping.
The hijacked RNode $E$ sabotages the instructions and continues to share two separate streams of Bell pair with $A$ and $B$.
$A$ and $B$ try to generate the secret key by measuring qubits they hold, but actually share the key with $E$, who is impersonating the other.
$E$ uses these keys to decrypt and re-encrypt the encrypted information transferred between $A$ and $B$.

\subsection{Attacks by a malicious router node}
An XNode is the most powerful QNode; the set of attacks it can perform are a superset of the attacks that an RNode can.
When an attacker aims to maximize the hijacking time and the range of influence while remaining undetected, the following attack means are available in addition to the above attacks~\cite{satoh2018network}.
\subsubsection{Switching disruption}
An attacker can execute entanglement swapping without following the distributed ruleset created to govern a connection~\cite{matsuo2019quantum}, and forward the Qubit to a node far away from the destination~(Fig.~\ref{fig:disswitch}).
We can detect the intentionally wrong entanglement swapping operation by Bell test, but we cannot avoid the negative effects on \Integ and \Avail before the detection~\cite{satoh2018network}.
\Figure[htbp][width=240pt]{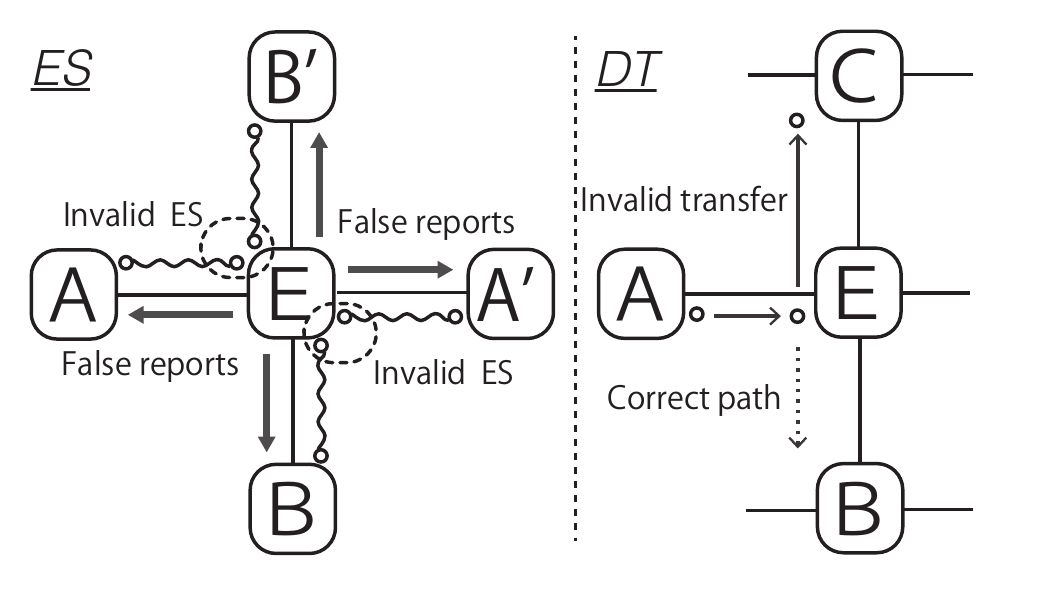}
{In an ES network, an attacker controlling hijacked Xnode E can share Bell pairs between incorrect node pairs $A-B'$ and $A'-B$ instead of correct pairs $A-A'$ and $B-B'$ using malicious entanglement swapping. On a DT repeater network, the attacker may intentionally transfer qubits to a inappropriate QNodes C instead of destination B. These disruptions decrease network \Integ and \Avail of transferred information.  \label{fig:disswitch}}

\subsubsection{Framing innocent repeaters}
\begin{figure}[htbp]
 \begin{center}
  \includegraphics[width=240pt]{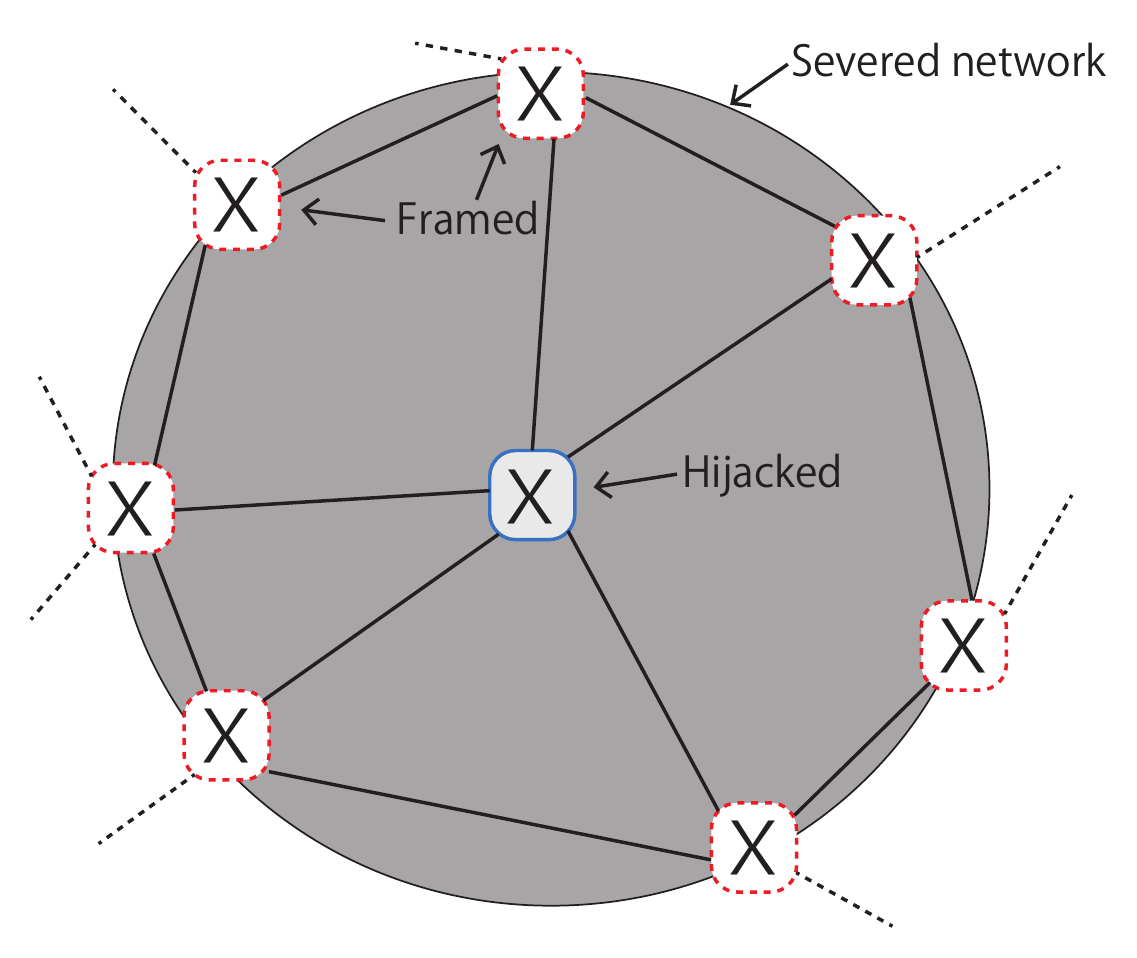}
  \caption{A network partitioned by the isolation of innocent QNodes. Red nodes denote isolated innocent nodes. The blue node denotes hijacked XNode. Solid lines denote working links. Dashed lines denote links cut by surrounding innocent nodes due to framing.
The repeated success of framing leads to this situation.
Due to differences in the frequency of communications, the closer a QNode is to a hijacked XNode, the more vulnerable it is.}
  \label{fig:framing}
 \end{center}
\end{figure}

The Quantum Internet will perform certification to monitor network quality and intrusion.
Malicious routers can ``frame'' other repeaters or routers as having failed or been hijacked by misreporting certification results or selectively interfering with the communication process through a particular node~\cite{satoh2018network}.
This framing can cause a repeater to be excluded from acting on the network. 
By framing a chosen set of routers (a \emph{separator}, in graph theory terms~\cite{doyle2005robust}), the malicious node can partition the network (Fig.~\ref{fig:framing}).
To minimize such threats, we need to detect and respond as soon as possible.

The loss of \Avail due to the isolation of a router from the network is more significant than other nodes due to the large amount of connectivity involved.
To avoid a significant network performance loss due to the isolation of a router, we should
keep the certification cycle secret, secure resources that do not identify the communication path and design the network with as rich a topology as possible.

\subsubsection{Path Black Hole}
Early on, and perhaps even into the indefinite future, the Quantum Internet will execute path setup first in the classical (management) plane, then quantum plane will start to generate End-to-End entanglement.
Therefore, the classical {\bf packet black hole attack} by advertising an incorrect address block results in collecting path setup packets, preventing the classical packets from arriving at the destination.  While this disrupts communication involving affected nodes or networks, because it does not result in new traffic on the quantum plane, any communications not involving the black holed nodes should continue unaffected.
This is different from dishonest quantum communication and attacks such as entanglement swapping sabotage by a single node, which result in wasting quantum plane resources.
\subsection{Attacks by multiple hijacked QNodes}\label{subsec:qn}
Next, we discuss the case where the attacker has successfully hijacked multiple QNodes.
The attack methods available to each node remain the same, but combining them will change the situation.
We investigate what kinds of attacks are possible or enhanced, depending on the combination of malicious nodes.

\subsubsection{Quantum distributed Denial-of-Service (QDDoS) attack}
First, we consider the possible attack methods for cooperating malicious ENodes and MNodes.
A distributed Denial-of-Service~(DDoS) attack is a method that compromises \Avail by attacking a single Internet service from many machines at once.
As in the classic DDoS, the DDoS in the Quantum Internet~(QDDoS) attacks available to end-users will be cost-effective and malicious.
We can divide QDDoS into attacks on the {\bf classical plane} as system controller and assaults on the {\bf quantum plane} as quantum resources provider.

The effect of the attack on the classical plane spills over to the quantum plane.
As with server crashes caused by DDoS, failure of the system controller can cause not only service outages but also loss of information on the terminal Qubit and QNIC Qubit.
The reconstruction of quantum information is more complicated than classical information.
The classic system responsible for managing these Qubits should be independent of the systems affected by QDDoS attack.

As a way to attack the quantum plane, we can expect an excessive number of service requests.
If the applications we provide are quantum key generation or cloud quantum computation, increasing the number of QNodes can address the attack.
If an application requires manipulation of certain quantum information, the attack can cause significant service delays~(\Avail) and information loss (\Integ).

QDDoS attack will be especially effective to serious in Quantum Internet because ENodes will have bandwidth.

\subsubsection{Framing using multiple hijacked QNodes}
In order for a hijacker to perform framing, the target QNode needs to be on the communication path.
Assuming no bias in network structure or frequency of use, the hijacker is less likely to frame QNodes that are farther away from the hijacked QNode.
If you succeed in multiple hijackings, the situation will change.
It is difficult to quantitatively discuss threats without defining the network structure, but
false reports from multiple nodes could more easily fool the network.

\section{Conclusion}\label{sec:conclusion}
This work is the first attempt to summarize the threats on the Quantum Internet. Modeling threats is an essential step toward providing countermeasures against attacks, and is essential to achieve secure and sustainable quantum networks.

We have provided an analysis of security for a quantum repeater architecture based on our current knowledge, by
referring to proposed taxonomies for classical systems.
By providing a model of a quantum
repeater network and grouping the elements of the modeled repeater, we provide a first look at the kinds of attacks
that may be possible.

From the point of view of \Conf, quantum repeater systems have great advantages.  Since it is possible to
detect the presence of {\bf an eavesdropper}, detection of a breach of \Conf is possible.  Quantum tomography sacrifices a
portion of our stream of Bell pairs as part of ongoing network monitoring operations as needed to tune certain physical
parameters to optimize the fidelity of our entanglement.  This process is extended to include eavesdropper detection by
choosing the portion sacrificed for tomography at random.  As long as tomography indicates that high fidelity is
achieved on the end-to-end connection, our remaining stream of entangled qubits can be safely used without fear of
breach of \Conf if the other end point and application are secure.

From the point of view of \Integ and \Avail, a quantum repeater system shares many of the vulnerabilities with a classical network system.
A repeater includes classical computing hardware and threats to both
\Integ and \Avail can target that hardware.

One of the keys to security of the quantum repeater system is not a quantum system specific issue, but rather the
classical parts of the system, including the classical part of the quantum node and classical network
services in the node, which are no different from classical network
equipment.  Mixed attacks making use of a combination
of quantum and classical parts may also prove to be an important topic.

One big difference is that quantum mechanics has the no-cloning theorem; 
quantum information cannot be copied to defend against loss in the network like classical networking.
In this sense, directly transmitting data qubits in the DT repeater network has more serious risk against DoS attacks.
This problem can be avoided by sending half Bell pairs even in the Third generation.
In that case, distributed connection management is required.
There will be choices of protocols, rapid and memory-efficient but risky protocol which sends data qubits directly,
and safe but slow and memory-inefficient protocol which sends half Bell pairs then executes quantum teleportation.

This paper, comprising a framework of attack points and goals, represents only the first step in assessing the security
of quantum networks.  We plan to extend our study further as engineers working in both classical and quantum networking,
to apply the lessons learned in classical networks to develop a full taxonomy of attacks, assess mitigation strategies,
and ultimately minimize security issues with developing quantum networks.

\section*{Acknowledgment}
The authors acknowledge members of the Quantum Internet Task Force (QITF), a research consortium working to realize the Quantum Internet, and participants in the Quantum Internet Research Group (QIRG) of the Internet Research Task Force, for comprehensive and interdisciplinary discussions of the Quantum Internet.

\bibliographystyle{IEEEtran}
\bibliography{satoh_quantum}
\EOD
\end{document}